\documentclass[a4paper]{article}

\usepackage[margin=1in]{geometry}
\usepackage[utf8]{inputenc}
\usepackage[T1]{fontenc}
\usepackage{amsmath}
\usepackage{microtype}
\usepackage{hyperref}
\usepackage{cleveref}

\usepackage[citestyle=authoryear,bibstyle=authoryear,backend=bibtex,natbib]{biblatex}
\bibliography{anti-ulanowicz}

\begin{document}


\title{Comments on ``the return of information theory''}
\author{Baltasar Trancón y Widemann \qquad Christina Bogner\\[\medskipamount]Ecological Modelling\\University of Bayreuth, Germany}

\date{Submitted to {\em Ecological Complexity}, 2012\\Rejected without review}

\maketitle

\begin{flushright}
  \begin{minipage}{0.4\linewidth}
    \textsl{You have found the Ulanowicz et al.\ paper inconsistent,
    misleading and terminologically incorrect. This is very bad, I
    agree.}

  \rightline{\footnotesize S.\ Petrovskii, {\em Ecol.\ Compl.} Editor-in-Chief}
  \end{minipage}
\end{flushright}
  

  


\section{Introduction}

In the course of our own research on theoretical concepts for ecology,
inspired by computing and information science, we have discovered the
recent\footnote{At the time of writing, 2012.} article ``Quantifying
sustainability: Resilience, efficiency and the return of information
theory'' \parencite{Ulanowicz2009}.  It has been a sincere pleasure to
discover that renowned ecologists keep advocating the use of
information theory for the description and assessment of ecosystems.
It has been a severe disappointment, however, that information theory
is far from being portrayed in the best possible light in the article
in question---the formulation and use of basic IT concepts are
unnecessarily obscure, deliberately incompatible with standard
terminology, and in places just mathematically wrong.

We shall endeavor to lift some of the confusion by correcting errors,
dissecting ambiguities and using standard terminology wherever
possible.  We believe that clear and rigorous presentations are
necessary in order to enable information theoreticians and ecologists
to engage in a fruitful dialogue.  Only very basic concepts of IT are
required to follow our arguments.  The interested reader is referred
to the textbook of \cite{MacKay2003}, an excellent and encyclopaedic
treatise that is freely accessible online.

This critique should not be understood as extending to the
\emph{content} of the criticized article; it adresses its formal
\emph{presentation} only.  To the contrary, our disappointment arises
mainly from fear that the article might influence readers rather
against than in favour of application of IT in ecology, because of the
vague air of inconsistency owing to the many terminological and
mathematical issues we are about to discuss.

\section{Terms}

\subsection{Surprisal and Shannon Information Content}

What \citeauthor{Ulanowicz2009} call ``Boltzmann's famous definition
of surprisal'', $s = - k \log p$, is nowadays known in IT as the
\emph{Shannon information content} of an event.  While Boltzmann
preceded Shannon by some 80 years, it was the latter who generalized
the idea, brilliant as it may have been, into a full-fledged theory.

The scalar constant $k$ is not explicit in the modern definition;
rather, it is implicit in the dimensionless unit implied by the choice
of base of the logarithm; cf.\ \cref{log-units} below.  The particular
constant $k_{\mathrm{B}}$ that nowadays bears Boltzmann's name is for
the particular case of thermodynamic entropy, and carries the
dimension of a heat capacity.
Shannon argues that, in IT, no single appropriate value exists, and
the constant ``merely amounts to a choice of a unit of
measure'' \parencite{Shannon1948}, in the sense that it is the only
degree of freedom in an axiomatic specification of the entropy
function, see below.

It is ironic that \citeauthor{Ulanowicz2009} elaborate on the sign of
the expression,
\begin{quote}
  ``Because the probability, $p$, is normalized to a fraction between
  zero and one, most offhandedly conclude that the negative sign is a
  mathematical convenience to make $s$ work out positive (and that may
  have been Boltzmann's motivation).  But from the perspective of
  logic {\bfseries one can only read this equation} as defining $s$ to
  gauge what $p$ is not.'' {\upshape[Emphasis added]}
\end{quote}
seeing that the form inscribed on Boltzmann's tombstone, $S = k \log
W$, has a positive sign instead.  Confer also the notation of
\cite{MacKay2003}, where the likewise positive form $\log_2 (1/P(x))$
is used.  Reciprocal pairs of quantities abound in science, for
instance consider ``period'' and ``frequency'', or ``resistance'' and
``conductance''.  Which of each pair is about what is and what is not,
respectively, depends very much on the frame of reference.

\subsection{Aggregate Indeterminacy and Entropy}

What \citeauthor{Ulanowicz2009} call the ``aggregate systems
indeterminacy'', defined in their equation (3) up to a missing
equality sign (a simple typographic error that may subtly confuse
the reader nevertheless), is nowadays usually called the \emph{Shannon
  entropy} of a random variable whose possible outcomes are specified
by the events $i$ and respective probabilities $p_i$.
\begin{equation*}
  H = - \sum_i p_i \log p_i
\end{equation*}
Because all of the following definitions involve more than one random
variable, it greatly adds to clarity to mention them explicitly:
\begin{equation*}
  H(Y) = - \sum_i P(Y=i) \log P(Y=i)
\end{equation*}
We shall use only the symbols $Y$ and $Z$ for random variables,
because $X$ is reserved in the notation of \cite{Ulanowicz2009}.

\subsection{Average Mutual Constraint and Mutual Information}
\label{mutual}

What \citeauthor{Ulanowicz2009} call the ``average mutual constraint''
(without saying precisely between what, see \cref{event-random,symm}),
defined in their equation (5), is usually called the \emph{mutual
  information} or \emph{transinformation} between two random
variables.
\begin{equation*}
  I(Y, Z) = \sum_{i,j} P(Y = i, Z = j) \log \frac{P(Y = i, Z = j)}{P(Y = i)P(Z = j)}
\end{equation*}
Note that the explanations leading to equation (5) are problematic in
various ways, see \cref{event-random,joint,symm}.

\subsection{Conditional Entropy}

What \citeauthor{Ulanowicz2009} call the ``conditional entropy''
$\Psi$ is \emph{not} what is usually called a conditional entropy.  A
typical definition of conditional entropy, or \emph{quivocation}, of
$Y$ given $Z$ would look like:
\begin{equation*}
  H(Y \mid Z) = \sum_{i,j} P(Y = i, Z = j) \log \frac{P(Z = j)}{P(Y = i, Z = j)}
\end{equation*}
It plays an important role in the decomposition of uncertainties in
IT, such as in the so-called \emph{chain rule}:
\begin{equation*}
  H(Y, Z) = H(Y \mid Z) + H(Z)
\end{equation*}
The definition of \citeauthor{Ulanowicz2009}'s equation (7) can be
retrieved by \emph{adding} the conditional entropies of $Y$ given $Z$ and
vice versa:
\begin{align*}
  H(Y \mid Z) + H(Z \mid Y) &= \sum_{i,j} P(Y = i, Z = j) \log
  \frac{P(Z = j)}{P(Y = i, Z = j)} +
  \\
  &\mathrel{\hphantom{=}} \sum_{i,j} P(Y = i, Z = j) \log \frac{P(Y =
    i)}{P(Y = i, Z = j)}
  \\
  &= \sum_{i,j} P(Y = i, Z = j) \log \frac{P(Y = i)P(Z = j)}{P(Y = i,
    Z = j)^2}
\end{align*}

This sum is sometimes called the \emph{variation of information}
between $Y$ and $Z$.  It is easy to see that it satisfies the axioms
of a \emph{metric} and may hence serve as an information-theoretic
measure of distance between random variables.

\section{Ambiguities and Errors}

\subsection{Events and Random Variables}
\label{event-random}

In equation (3), \citeauthor{Ulanowicz2009} give the (slightly broken)
definition of the Shannon entropy of a single random variable.
Immediately afterwards, they move on to the more complex topic of a
pair of random variables, but without explicitly saying so.  The
transition is implied in the statement:
\begin{quote}
  ``Accordingly, we will define $p_{ij}$ as the joint probability that
  events $i$ and $j$ co-occur.''
\end{quote}
The failure to explicitly identify the two random variables, only
distinguished by the fact that their outcomes are indexed with
variables $i$ and $j$, respectively, causes confusion even with the
authors themselves: They apparently fail to realize that what they
call \emph{the} conditional entropy is in fact the sum of two
conditional entropies, namely of each random variable conditional on
the other.  Consequently, the interesting question whether the two
should be lumped together or examined separately is not addressed: For
instance, calculation of conditional entropies of $Y$ (origin of flow)
given $Z$ (destination of flow) and vice versa for the networks
depicted in \citeauthor{Ulanowicz2009}'s Figs.~1 and 3 are strongly
asymmetric, with ratios of $1:6$ and $1:8.6$, respectively.  We do not
have an ecological rationale why this should be meaningful, but we
feel that a distinction that arises naturally from standard
terminology should not be obscured deliberately, unless it has been
proven irrelevant.

The same source of imprecision might also have contributed to the
problem discussed in the next subsection.

\subsection{Joint Indeterminacy and Independency}
\label{joint}

In their explanation of ``average mutual constraint'' (mutual
information), between their two equations both numbered (4),
\citeauthor{Ulanowicz2009} state:
\begin{quote}
  ``Here the assumption is made that the indeterminacy $s_{ij}$
  {\upshape [$= -k\log(p_{ij})$]} is maximal when $i$ and $j$ are
  totally independent.  We call that maximum $s_{ij}^*$.''
\end{quote}
The assumption does not hold, and there is no such maximum: For
independent events, we have $p_{ij} = p_{i.} p_{.j}$, but joint
probabilities can of course be smaller than that value.  In
particular, $s_{ij} = +\infty$ for mutually exclusive events. 

Stating that $s_{ij}$ is bounded above by $s_{ij}^*$ is equivalent to
stating that $p_{ij}$ is bounded below by $p_{i.}p_{.j}$.  We can only
conjecture that this (false) assumption is made in order to ensure
that mutual information is nonnegative.  Fortunately, the assumption
is not needed: Individual terms $\log (p_{ij}/p_{i.}p_{.j})$ in
equation (5) may well be negative, but their weighted sum, the mutual
information, is nonnegative by Gibbs' inequality; see also
\cref{log-role}.

\subsection{Symmetry of Mutual Information}
\label{symm}

\citeauthor{Ulanowicz2009} claim that their second equation (4) is
symmetric, written as $x_{i|j} = \dots = x_{j|i}$.  Note that this is
\emph{not} symmetry in the usual algebraic sense, namely that the
indices $i$ and $j$ may simply be exchanged.  If one does so
na{\"\i}vely, the denominator of the argument to the logarithm becomes
$p_{j.}p_{.i}$ which is very different from $p_{i.}p_{.j}$.  On the
other hand, if one remembers that $i$ and $j$ are events concerning
two different random variables $Y$ and $Z$, respectively, then the
symmetry becomes obvious, as both $P(Y = i, Z = j) = P(Z = j, Y = i)$
and $P(Y = i)P(Z = j) = P(Z = j)P(Y = i)$ hold trivially. Hence it
follows that $I(Y, Z) = I(Z, Y)$; see \cref{mutual} above.

\subsection{The role of the Logarithm}
\label{log-role}

Much has been said about the role of the logarithm in the formulae of
IT. \citeauthor{Ulanowicz2009} promise in their footnote~1:
\begin{quote}
  ``Here the reader might ask why the lack of $i$ is not represented
  more directly by $(1 - p_i)$? The advantage and necessity of using
  the logarithm will become apparent presently.''
\end{quote}
But only a very brief rationale can be found in the following
discussion, qualifying hardly as advantageous, and certainly not as
necessary: In the statement leading to their equation (6) it is
claimed (correctly) that the convexity of the logarithm ensures that
joint entropy decomposes into mutual information and conditional
entropies, all non-negative.  But that would be true for any other
convex function.  \cite{Shannon1948} has the definitive answer to
the riddle: Any function satisfying a small number of properties
characteristic of a measure of uncertainty (namely continuity,
monotonicity in the number of outcomes of uniform distributions, and
additive compositionality of choice) is necessarily equivalent to
Shannon entropy, up to a conversion of units.

Instead of actually working with logarithms,
\citeauthor{Ulanowicz2009} promptly revert to non-logarithmic scales
in their section~5, in the introduction to the concept of ``window of
vitality''.  There they cite previous work \parencite{Zorach2003},
where the measures are developed in entirely non-logarithmic form.
For additional serious problems with the concerned paragraphs, see
\cref{expo} below.

\subsection{Logarithmic Units}
\label{log-units}

In the argument leading towards their definitions of
``ascendency'' and ``reserve'', after equation (10),
\citeauthor{Ulanowicz2009} state:
\begin{quote}
  ``The dimensions in the definitions (10) remain problematic,
  however. All of the ratios that occur there are dimensionless (as
  required of probabilities), so that the only dimensions that the
  variables $H$, $X$ and $\Psi$ carry are those of the base of the
  logarithm used in their calculation.  For example, if the base of
  the logarithm is 2, the variables are all measured in bits.''
\end{quote}
In this statement, the concepts of \emph{unit} and \emph{dimension}
are confused.  A unit conveys two independent aspects of meaning:
dimension and \emph{magnitude}.  For example, the SI unit
$1\,\mathrm{m}$ has the dimension \emph{length} and a magnitude in
relation to other units of length that makes it equivalent to, say,
$39.37\,\mathrm{in}$.  The units of information, just like other
logarithmic quantities, are \emph{dimensionless} units, but they do
have a magnitude.  For example, $1\,\mathrm{bit}$ is equivalent to
$1/8\,\mathrm{byte}$, about $1.44\,\mathrm{nat}$ or
$3.01\,\mathrm{dB}$. If we take the last sentence of the above
quotation as implying that the base of the logarithm is not fixed once
and for all, then the magnitude of the unit of IT measures carries
essential meaning.

In the immediately following section~4, ``A two-tendency world'',
\citeauthor{Ulanowicz2009} give concrete numbers for material flows in
an ecosystem and purport to multiply IT measures with total flow:
\begin{quote}
  ``$T_{..}$ for this system is
  $102.6\,\mathrm{mg}\,\mathrm{C}\,\mathrm{m}^{-2}\,\mathrm{y}^{-1}$;
  the ascendency, $A$, works out to
  $53.9\,\mathrm{mg}\,\mathrm{C}\,\mathrm{bits}\,\mathrm{m}^{-2}\,\mathrm{y}^{-1}$
  and the reserve, $\Phi$, is
  $121.3$ $\mathrm{mg}\,\mathrm{C}\,\mathrm{bits}\,\mathrm{m}^{-2}\,\mathrm{y}^{-1}$.''
\end{quote}
Have you noticed the difference in units between the first and the
following figures?  That they are given in one sentence creates the
dangerously false impression that the absolute values can be compared
as ``apples with apples''
\parencite[section 7, ``One-Eyed Ecology'']{Ulanowicz2009}.  The
practice is even worse than ambiguous: the use of several decimal
places and fully explicit dimensional units suggest an absoluteness
that is not warranted. The above figure for $A$ could be given as
either
\begin{itemize}
\item
  $53.9~\mathrm{mg}\,\mathrm{C}\,\mathrm{bit}\,\mathrm{m}^{-2}\,\mathrm{y}^{-1}$,
  in the logarithmic units of base $2$ chosen by the authors, or
\item
  $37.7~\mathrm{mg}\,\mathrm{C}\,\mathrm{nat}\,\mathrm{m}^{-2}\,\mathrm{y}^{-1}$,
  in the logarithmic units of base $e$ used by Boltzmann, or
\item
  $162.3~\mathrm{mg}\,\mathrm{C}\,\mathrm{dB}\,\mathrm{m}^{-2}\,\mathrm{y}^{-1}$,
  in the logarithmic units of base $10^{1/10}$ preferred by electric
  engineers, or
\item
  $6.74~\mathrm{mg}\,\mathrm{C}\,\mathrm{byte}\,\mathrm{m}^{-2}\,\mathrm{y}^{-1}$,
  in the logarithmic units of base $2^8$ (such that $1~\mathrm{byte} =
  8~\mathrm{bit}$) preferred by many computer scientists, or
\item
  $9.88~\mathrm{pg}\,\mathrm{C}\,\mathrm{CDROM}\,\mathrm{m}^{-2}\,\mathrm{y}^{-1}$,
  in the logarithmic units corresponding to the information content
  of a standard (Red Book) Mode-1 CD-ROM of $666\,000 \times
  1024~\mathrm{byte}$, or
\item
  $64.7~\mathrm{g}\,\mathrm{C}\,\mathrm{cent}\,\mathrm{m}^{-2}\,\mathrm{y}^{-1}$,
  in the logarithmic units of base $2^{1/1200}$ preferred by musicians,
\end{itemize}
or in whatever logarithmic units another author might fancy.

Having ruled out the possibility to compare flows with
flow--information products directly, and having conceded that
flow--information products only make sense when compared to each other
in appropriately adjusted units, one wonders why the IT measures have
been multiplied with flows in the first place; all of the arguments of
their section~4 would have worked perfectly, and with less room for
confusion, with $X$ and $\Psi$ in place of $A$ and $\Phi$.  See also
the next subsection.

\subsection{Exponential Units in the Window of Vitality}
\label{expo}

Even the reader who has worked through the bookkeeping of dimensional
and dimensionless units in the section we have just discussed, must be
baffled by the following statement \cite[section~5]{Ulanowicz2009},
``The survival of the most robust'':
\begin{quote}
  ``Zorach and Ulanowicz (2003) [\dots] plotted the networks, not on
  the axes $A$ vs.\ $\Phi$, but rather on the transformed axes $c =
  2^{\Phi/2}$ and $n = 2^A$.
\end{quote}

Certainly this cannot be correct, because here dimensional quantities
appear in the exponents.  Cross-reading of the cited
source \parencite{Zorach2003} reveals a definition of the symbol
$\Phi$ equivalent to the definition of the symbol $\Psi$ by
\cite{Ulanowicz2009}. Although more difficult to verify, we conjecture
that $A$ should be read as $X$ accordingly.  That is, after attaching
physical dimensions to information measures by multiplying them with
flows, the authors silently revert to the dimensionless quantities, in
order to make exponential scaling meaningful.

What makes the comparison of information measures for different
systems possible is exactly what is criticized scornfully in their
section~3:
\begin{quote}
  ``Unfortunately, bits do not convey any sense of the physical
  magnitude of the systems to which they pertain. For example, a
  network of flows among the populations of microbes in a Petri Dish
  could conceivably yield an $H$ of the same order of magnitude as a
  network of trophic exchanges among the mammalian species on the
  Serengeti Plain.''
\end{quote}
This feature of IT, namely that it quantifies information content of
observed choice (traditionally of messages in communication) as
\emph{intensive} properties, regardless of the extent of systems that
produce them, is commonly regarded as one of its essential
abstractions.  Nor is this an uncommon practice in science; certainly
no one would object to the temperature of a Petri Dish being compared
to the temperature of the Serengeti Plain.  Temperature as an example
has not been chosen randomly; indeed Boltzmann's work on uncertainty
has had its main application in thermodynamics.

\section{Conclusion}

Interdisciplinary research is partially motivated by the hope that
innovative research can be sparked by exposing experts of one field to
the concepts of another.  The article criticized here, though clearly
intended for the very purpose, is an illustrative example how
\emph{not} to do it.  An overview article in a journal is not the
place to provide a sound and complete introduction to a theory, but it
can and should leave a non-expert reader both motivated and prepared
to read up on the details in the literature of the field.

We have identified two key issues that merit words of warning and
recommendations to prospective interdisciplinary researchers:
\begin{enumerate}
\item \emph{Ideosyncratic terminology.} Renaming key concepts of
  theories in the process of translating them to a different field
  leaves the reader unfit to look up the technical details in the
  relevant literature.  Theoreticians have a cause for celebration
  whenever they discover that concepts in disparate fields are
  actually the same; actively propagating obfuscated terminology
  hinders theoretical progress and may well cause the reinvention of
  several wheels.  
\item \emph{Internal inconsistency.}  A reader who is not an expert on
  the subject may not fully grasp, or explicitly doubt, the results of
  a mathematical discussion at first sight; scepticism is generally a
  laudable trait for a scientist.  However, there are numerous
  heuristics to judge whether the arguments in question can possibly
  be valid: Mathematical digressions should be both true and relevant,
  dimensions and units should match for all figures to be compared,
  all mathematical objects involved in a definition should be named,
  etcetera.  All maths, especially when given to demonstrate the
  usefulness of a theory, should be prepared carefully enough to pass
  the test of heuristic reading by a reasonably educated
  non-expert.  On the other side, editors should feel encouraged to
  seek reviews of interdisciplinary articles from experts in the
  respective other field.  Thus, issues as the ones we have pointed
  out above, and possibly less obvious ones also, could be remedied
  before publication.
\end{enumerate}

A description of an unfamiliar theory in terms that cannot be traced
to their usage in the field, and in formulae that obviously cannot be
quite right, makes a tough reading.  Who can blame the readers for
concluding that the subject is not worth their attention?  However,
that conclusion could be severely premature.

In the present case, we see no reason to doubt the validity or the
utility of \citeauthor{Ulanowicz2009}'s philosophical observations and
empirical findings.  The proposal to investigate the \emph{potential}
(as opposed to \emph{actual} as in philosophy, not as opposed to
\emph{kinetic} as in physics) is certainly sound.  We conjecture that
similar investigations would be worthwhile for other subdisciplines of
what is nowadays subsumed under computing and information science, for
instance in modal, fuzzy and other nonclassical logics, or
nondeterministic automata theory.  Evidence of tentative import into
ecology can be found in the logical study of
uncertainty \parencite{Regan2002}, and in the use of Petri nets (not
to be confused with Petri dishes) for ecological modelling by
\cite{Sharov1991}, respectively.  The latter is a fine case in point
of our critique, as an interdisciplinary application that is also
quite up to the internal standards of the Petri net community.  We are
looking forward to an ecological presentation of information measures
of comparable quality.

\printbibliography

\end{document}